\documentclass{article}

\usepackage{amsmath}
\usepackage{cite}
\begin{document}
\begin{titlepage}
\noindent{\large \bf Cluster approximation solution of a two species annihilation model}

\vskip 2 cm

\begin{center}{F. Tabatabaee$^{a,}${\footnote
{ftabatabaee@hotmail.com}} , A. Aghamohammadi$^{a,b,}${\footnote
{mohamadi@alzahra.ac.ir}} } \vskip 5 mm

{\it  $^a$ Department of Physics, Alzahra University,
             Tehran 19834, Iran. }

{\it  $^b$ Institute for Applied Physics, Tehran 15857-5878 Iran.}
\end{center}

\begin{abstract}
\noindent
A  two species reaction-diffusion model, in which particles diffuse on
a one-dimensional lattice and annihilate when meeting each other,
has been investigated. Mean field equations for general choice of
reaction rates have been solved exactly.
Cluster mean field approximation of the model is also studied.
It is shown that, the general form of large time behavior of
one- and two-point functions of the number operators, are determined
by the diffusion rates of the two type of species, and is independent
of annihilation rates.
\end{abstract}
\begin{center} {\bf PACS numbers:}

{\bf Keywords:} reaction-diffusion, cluster mean field
\end{center}

\end{titlepage}
\newpage
\section{Introduction}
Recently properties of systems far from equilibrium have been
studied by many people. Although mean field techniques may give
qualitatively correct results for higher dimensions, for
low-dimensional systems fluctuations have important roles.
Different methods have been used to study reaction-diffusion
models, including analytical and approximational methods. Among
them, there are methods to obtain some quantities can be obtained
exactly. For example in \cite{Gs,AAMS,AM}, imposing some
constraints on the reaction rates leads to a closed set of
equations for average number densities in each site. The empty
interval method, is another method, which has been also used to
analyze the one dimensional dynamics of diffusion-limited
coalescence \cite{BDb,BDb1,BDb2,BDb3}. The most general
one-dimensional reaction-diffusion model with nearest-neighbor
interactions that can be solved exactly through empty interval
method, has been introduced in \cite{AKA}. Empty interval method
has been also generalized in \cite{Mob,AAK}. Different methods has
been introduced to calculate different quantities exactly.
However, exactly solvable models are only a special class of
reaction-diffusion models, and so people are motivated to use also
approximate methods to understand the role played by fluctuations.
In \cite{MH} a two species model has been considered. In this
model there are three competing reactions $AA\to \emptyset$,
$BB\to \emptyset$, and $AB\to \emptyset$. Asymptotic density decay
rates of the two type of species for a special choice of
parameters have been studied using the Smoluchowski approximation
and also field theoretic renormalization group techniques. A
similar model focusing on the same diffusion rates for the two
type of species has been studied in \cite{KJ}. Field theoretic
renormalization group analysis suggest that contrary to the
ordinary mean-field technique, the large time density of the
minority species decays at the same rate as the majority ones in
one-dimensional case. Although ordinary mean-field technique,
generally do not give correct results for low-dimensional systems,
its generalizations such as cluster mean-field may give correct
results. Any how, in the mean field approximation at most
one-point functions may be obtained. To obtain more-point
functions one should use other methods. One possible way is, to
use a generalization of mean field known as the cluster mean field
approximation.

One of the topics, which have absorbed many interests in recent
years, is non-equilibrium phase transitions. There are examples,
in which mean field (MF) solutions are not valid, but its
generalization, cluster mean field (CMF) gives qualitatively
correct results \cite{Od,MSM,DRO,ODS}. A coagulation-production
model is recently considered in \cite{Od}. Although MF equations
do not give correct results, CMF approximation predicts phase
transition, supported also by Monte Carlo simulations. Steady
state properties in the absorbing phase of 1{\bf d} pair contact
process model are also investigated using Monte Carlo simulations
and the cluster approximation. The cluster approximation
qualitatively confirms the numerical results \cite{MSM}.

The scheme of the paper is as follows. In section 2, The mean field
equations for general  parameters have been solved exactly. It is seen
that, the large time behavior of the average densities depend both on
initial average densities and reaction rates, and are independent of
diffusion rates. In section 3, The cluster mean field equations for
one- and two-point functions have been solved numerically. It is shown
that the general large time behavior is determined  by the diffusion rates.

\section{The Mean Field Approximation}
The model addressed in this
article is a two-species exclusion reaction-diffusion model. That
is, each site is a vacancy ($\emptyset$)  or at most occupied by a particle $A$
or $B$. The interaction is between nearest sites, and the adjacent
sites interact according to the following interactions with the
indicated rates.
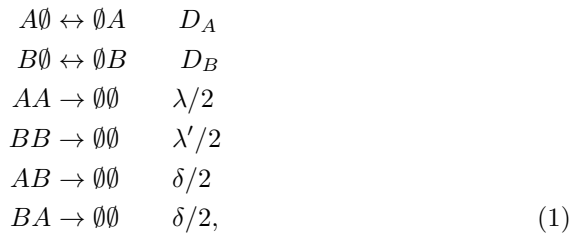
\begin{align}\label{1}
 A\emptyset&\leftrightarrow \emptyset A \qquad D_A\nonumber \\
 B\emptyset&\leftrightarrow \emptyset B \qquad  D_B\nonumber \\
 AA&\to\emptyset \emptyset \qquad  \lambda/2\nonumber \\
 BB&\to \emptyset \emptyset \qquad \lambda'/2\nonumber \\
 AB&\to \emptyset \emptyset \qquad  \delta/2\nonumber \\
 BA&\to \emptyset \emptyset \qquad  \delta/2,
\end{align}
We consider translationally invariant initial conditions. In
the mean-field approximation, diffusion rates do not have any
effect on the evolution equations of average number densities. The
mean-field equations for the average densities $a:=\langle A\rangle_t$ and
$b:=\langle B\rangle_t$ are
\begin{align}\label{2}
&{{\rm d}a\over {\rm d}t}=-\lambda a^2-\delta ab\nonumber \\
 &{{\rm d}b\over {\rm d}t}=-\lambda' b^2-\delta ab.
\end{align}
The large time behaviors of these  equations for special choices
of parameters have been studied in \cite{KJ,MH}. Now, we want to
solve these equations exactly and then we will show that there are
cases which are not considered in \cite{KJ,MH}, and give
qualitatively correct result for large time behaviors, although
the exponent of the decay rate is not correct.

Consider the following cases.

\vskip 1cm
\noindent I) $\lambda=\lambda'$.

\noindent The evolution equation for  $u:=b/a $, is
\begin{equation}\label{3}
{{\rm d}u\over {\rm d}t}=(\lambda-\delta) u(1-u)a.
\end{equation}
Using (\ref{2},\ref{3}), it is seen that
\begin{equation}\label{4}
{{\rm d}u\over {\rm d}a}=-{(\lambda-\delta) u(1-u)\over (\lambda
+\delta u)a},
\end{equation}
which can be integrated to
\begin{equation}\label{5}
  \left\vert{u-1\over u_0-1}\right\vert^{1+\delta/\lambda}({u_0\over u})=\left(
  {a\over a_0}\right)^{1-\delta/\lambda},
\end{equation}
where $u_0$ and $a_0$ are the initial values of $u$ and $a$,
respectively. Now we can obtain the large time behavior of the average
densities. It is seen that the large time behavior of $u$ depends
on the ratio $\delta/\lambda$.

\vskip 0.5cm
\noindent I.1) $\delta>\lambda$

\noindent At large times, obviously $a\to 0$, so it is seen from
(\ref{5}) that depending on the initial value $u_0$, two case may
occur

\noindent At large times $u\to \infty\qquad \Rightarrow\qquad
u\sim a^{\lambda/\delta -1},\quad b\sim a^{\lambda/\delta}$

\noindent At large times $u\to 0\qquad \Rightarrow\qquad u\sim
a^{\delta/\lambda  -1},\quad b\sim a^{\delta/\lambda}$

Assuming an imbalance in the initial average densities, for
example  $a_0>b_0$ ($u_0<1$), (\ref{2}) gives the large time
behavior of $a(t)$, and $u(t)$ as
\begin{align}\label{6}
&a(t)\sim t^{-1}\nonumber \\ & u(t)\sim t^{1-\delta/\lambda},
\end{align}
which means that for $\delta>\lambda$, in the mean-field
approximation the minority species dies out earlier than the majority
one, and the decay exponent of $u(t)$ is independent of diffusion
rates.

\vskip 0.5cm

\noindent I.2) $\delta<\lambda$

\noindent As a consequence of the large time behavior of $a$, $a\to
0$, it is seen from (\ref{5}), that at large times $u\to 1$.
Defining $\epsilon:=\mid 1-u\mid$,
\begin{equation}\label{7}
  \epsilon\sim a^{1-\delta/\lambda\over 1+\delta/\lambda}.
\end{equation}
To obtain the large time behavior of $a$ and $u$, we should use
again (\ref{2}, \ref{3}), which give
\begin{align}\label{8}
&a(t)\sim t^{-1}\nonumber \\ & \mid 1-u(t)\mid\sim
t^{-{1-\delta/\lambda\over 1+\delta/\lambda}},
\end{align}
which means that at large times  both the minority and the majority
species decay with the same rate. The exponent of decay rate does
not depend on the diffusion rates.

\vskip 1cm

\noindent II) $\lambda\ne\lambda'$.

\noindent For this case one arrives at
\begin{equation}\label{9}
{{\rm d}u\over {\rm d}a}=-{[(\lambda-\delta)-(\lambda'-\delta)
u]u\over (\lambda +\delta u)a},
\end{equation}
which after integration gives,
\begin{equation}\label{10}
 \left\vert {1-{(\lambda'-\delta)u\over \lambda-\delta}\over 1-
  {(\lambda'-\delta)u_0\over \lambda-\delta}}
 \right\vert^{(\delta^2-\lambda'\lambda)/(\delta-\lambda')}
  ({u\over u_0})^{-\lambda}=\left(
  {a\over a_0}\right)^{-\delta+\lambda}.
\end{equation}
Now, it is easy to obtain large time behavior of the average
densities. Generally, there are three cases,
\vskip 0.5cm
\noindent II.1) $\delta >\lambda, \lambda'$

\noindent Depending on the initial average densities, the large time
behavior of the average densities ratio is $u(t)\to 0$ ($b\sim
a^{\delta/\lambda}$) or $u\to \infty$ ($b\sim
a^{\lambda'/\delta}$), which means that one kind of species decays
faster.

\vskip 0.5cm
\noindent II.2) $\delta <\lambda, \lambda'$

\noindent Defining $\epsilon := u -{\lambda -\delta\over \lambda'
-\delta}$ , at large times $\epsilon(t)\sim
a^{[(\delta-\lambda)(\lambda' -\delta)]\over [\lambda
\lambda'-\delta^2]}$. In this case two kind of species decays with
the same rate.

\vskip 0.5cm
\noindent II.3) $\lambda<\delta <\lambda'$

\noindent At large times  the average densities ratio $u(t)\to 0$, and
$b\sim a^{\delta/\lambda}$.

\vskip 0.5cm
As it is seen, in the MF approximation only for a special
choice of parameters, which is independent of diffusion rates, the two
types of species decay with the same rate.
The case with $\lambda=\lambda'<\delta$, and $D_A=D_B$ has been considered in
\cite{KJ}. Using field theoretic renormalization group analysis,
it is shown that in one-dimension both type of species decay with
the same rate. Monte Carlo data also supports the field theory
predictions in the one-dimensional model.

\section{The Cluster Mean Field Approximation}
Now, we want to  use  cluster mean-field approximation. If the
diffusion rate for both type of species is the same, $N=2$ cluster
mean-field approximation gives the same value for the decay rates
for both type of species, even if there is an imbalance in the initial
average densities. If two type of species diffuse with
different rates, irrespective of the initial values, at large
times particles with greater diffusion rates  decay more rapidly.
For the nearest-neighbor interactions, the evolution equation of
$k$-point functions $\langle n_1n_2\cdots n_k\rangle$ contains at
most $(k+1)$-point functions. So, generally this set of equations
will be a hierarchy, which can not be solved exactly. One way to
overcome this difficulty is to impose constraints on the reaction
rates that leads to disappearance of $(k+1)$-point functions from
the evolution equation of $k$-point functions. This method has
been used to calculate some correlators  exactly in \cite{Gs}.
Another possible way is to use the cluster approximation. In the
$k$-site cluster approximation, the set of evolution equations
truncates and one encounters with a closed set of equations which
may be solvable, at least numerically. Any how, for a two-site
cluster approximation, a three site joint probability for a
sequence of nearest-neighbor sites is approximated by
\begin{equation}\label{12}
  P(A,B,C)=P(A \mid B,C)P(B,C)\simeq {P(A,B)P(B,C)\over P(B)}.
\end{equation}
where $P(A\mid B)$ is the conditional probability. In the
mean-field approximation there are three variables,
$\langle A\rangle$, $\langle B\rangle$,
and  $\langle\emptyset\rangle$,  only two of them are generally
independent. In the two-site cluster approximation, or pair
approximation, the variables
are $\langle A\rangle $, $\langle B\rangle$, $\langle\emptyset\rangle$,
 $\langle
A\emptyset\rangle$, $\langle B\emptyset\rangle$,
$\langle\emptyset\emptyset\rangle$, $\cdots$, among them there are
six independent variables which  we choose to be $\langle
A\rangle$, $\langle B\rangle$, $\langle AA\rangle$,$\langle
BB\rangle$, $\langle AB\rangle$, and $\langle BA\rangle$. In fact
in the pair approximation, besides the average densities the
two-point functions can also be obtained. The equation of motion
for the average densities when $D_A=D_b=:D$ are
\begin{align}\label{13}
&{{\rm d}\langle A\rangle\over {\rm d}t}=
-\lambda \langle AA\rangle-{\delta\over
2}\langle AB\rangle-{\delta\over 2}\langle BA\rangle\nonumber \\
&{{\rm d}
\langle B\rangle\over {\rm d}t}=-\lambda
\langle BB\rangle-{\delta\over 2}\langle AB\rangle-{\delta\over 2}
\langle BA\rangle.
\end{align}
Similar to the mean-field approximation, the diffusion rates do
not appear in the evolution equations of the average densities. But in
fact they affect the average densities through the evolution equations
of two-point functions, which are,
\begin{align}\label{14}
{{\rm d}\langle AA\rangle\over {\rm d}t}=&-{\lambda\over 2} \langle  AA
\rangle-\lambda
\langle AAA\rangle-{\delta\over 2}\langle AAB\rangle-{\delta\over 2}\langle
BAA\rangle\nonumber \\ &
-D\langle AA\emptyset\rangle-D\langle\emptyset AA\rangle+2D\langle A
\emptyset A\rangle,
\end{align}
\begin{align}\label{15}
{{\rm d}\langle BB\rangle\over {\rm d}t}=&-{\lambda\over 2} \langle BB\rangle-\lambda
\langle BBB\rangle-{\delta\over 2}\langle BBA\rangle-{\delta\over 2}\langle ABB\rangle
\nonumber \\ & -D\langle BB\emptyset\rangle-D\langle\emptyset BB\rangle+2D\langle
B\emptyset B\rangle,
\end{align}
\begin{align}\label{16}
{{\rm d}\langle AB\rangle\over {\rm d}t}=&-{\delta\over 2} \langle AB\rangle-
{\delta\over 2}
\langle  BAB\rangle-{\delta\over 2} \langle ABA\rangle-D\langle AB\emptyset
\rangle\nonumber \\ &-D\langle\emptyset
AB\rangle+2D\langle A\emptyset B\rangle,
\end{align}
\begin{align}\label{17}
{{\rm d}\langle BA\rangle\over {\rm d}t}=&-{\delta\over
2} \langle BA\rangle-{\delta\over 2} \langle BAB\rangle-{\delta\over 2}
\langle ABA\rangle-
D\langle BA\emptyset\rangle\nonumber \\ &-D\langle \emptyset BA\rangle+2D
\langle B\emptyset A\rangle.
\end{align}
To solve these equations in the cluster approximation, one should first
approximate three-point functions and then all the equations should be
expressed  in terms of independent variables. For example
$\langle AB\emptyset\rangle$ can be written as
\begin{equation}\label{18}
  \langle AB\emptyset\rangle\approx {\langle AB\rangle \langle B\emptyset\rangle
  \over \langle B\rangle}
\end{equation}
and then  using probability conservation  $\langle B\emptyset\rangle$
should be expanded,
\begin{equation}\label{19}
  \langle B\emptyset\rangle=\langle B\rangle-\langle BA\rangle-
  \langle BB\rangle .
\end{equation}

\subsection{ $D_A=D_B$}
Figure 1 and 2  show results for $\langle A\rangle$, $\langle B\rangle$, and
the density ratios $u(t):=\langle B\rangle/\langle A\rangle$ obtained using
numerical solutions of equations
(\ref{12}-\ref{16}). As it is seen both types of species decay with the
same rate irrespective of equality or inequality of reaction rates
$\lambda$ and $\lambda'$. In the MF approach, $K(t):=\langle BA\rangle/
\langle AA\rangle$is
not an independent quantity and is $\langle B\rangle/\langle A\rangle$.
But in the CMF
approach it is an independent one and the numerical result
obtained for it is plotted in figure 3. As it is seen, in the CMF
approximation it approaches a constant value at large times, means
that both $\langle AA\rangle$ and $\langle BA\rangle$ decay with the same
rate. Equality of
their decay rates is independent of equality or inequality of
reaction rates $\lambda$ and $ \lambda'$.

\subsection{ $D_A\ne D_B$ }
As MF equations are independent of diffusion rates, their
solutions remain unaltered. But in the pair approximation, only
equations (\ref{13}) remains unaltered. The diffusion rate $D$
in (\ref{14},\ref{15}) should be changed properly to $D_A$ or $D_B$,
and the equations (\ref{16},\ref{17}) become.

\begin{align}\label{20}
{{\rm d}\langle AB\rangle\over {\rm d}t}=&-{\delta\over 2}
\langle AB\rangle-{\delta\over 2}
\langle BAB\rangle-{\delta\over 2} \langle ABA\rangle-D_B\langle AB
\emptyset\rangle\nonumber \\ &-D_A\langle \emptyset
AB\rangle+(D_A+D_B)\langle A\emptyset B\rangle,
\end{align}
\begin{align}\label{21}
{{\rm d}\langle BA\rangle\over {\rm d}t}=&-{\delta\over
2} \langle BA\rangle-{\delta\over 2} \langle BAB\rangle-{\delta\over 2}
\langle ABA\rangle-
D_A\langle BA\emptyset\rangle\nonumber \\ &-D_B\langle \emptyset
BA\rangle+(D_A+D_B)\langle
B\emptyset A\rangle.
\end{align}
These set of equations has  been solved numerically, and the
numerical results for the average densities has been plotted in
figures 4 and 5. As it is seen, at large times, species with
greater diffusion rate dies out faster. If  species with greater
diffusion rate are majority initially, there is  a cross over, as
it is seen from figure 4. The general behavior of the average
densities ratio is independent of $\lambda$,  $ \lambda'$, and
$\delta$, and the general form of large time behavior is
determined by the diffusion rates. See figure 6. The numerical
results for $K(t)$ have been summarized in figure 7.

\vskip\baselineskip

\noindent {\bf Acknowledgement} \\ A. A. would like to thank M. Khorrami
for useful discussions.

\newpage
\noindent {\Large \bf{Figure captions}}
\vskip 1cm
\noindent figure 1- Average densities
$\langle A\rangle$, and $\langle B\rangle$ as a
function of time. The rates are $D_A=D_B=1$,
$\lambda=\lambda'=1000$, and $\delta =3000$.
\vskip 0.5cm
\noindent figure 2- Ratio of average densities,
$u(t)=\langle A\rangle /\langle B\rangle$,
as a function of time. For the dashed line, the rates are
$D_A=D_B=1$, $\lambda=1000$,
$\lambda '=500$,  and $\delta =3000$.
\vskip 0.5cm

\noindent figure 3-  $K(t)=\langle BA\rangle /\langle AA\rangle$,
as a function of time.
\vskip 0.5cm

\noindent figure 4- Average densities as
a function of time. The rates are $D_A=0.1$, $D_B=1$.
\vskip 0.5cm

\noindent figure 5- Average densities as a function of time. The rates are
$D_A=1$, $D_B=0.1$.

\noindent figure 6- $u(t)=\langle B\rangle /\langle A\rangle$,
as a function of time.
The rates are $\lambda=2400$, $ \lambda'=1000 $, $\delta =2500$,
and the greater
diffusion rate is $1$, and the smaller one, is $0.1$.
\vskip 0.5cm

\noindent figure 7- $K(t)=\langle BA\rangle /\langle AA\rangle$,
as a function of time.
The rates are $\lambda=250$, $ \lambda'=100 $, $\delta =300$,
and the greater diffusion rate
is $1$, and the smaller one, is $0.1$.

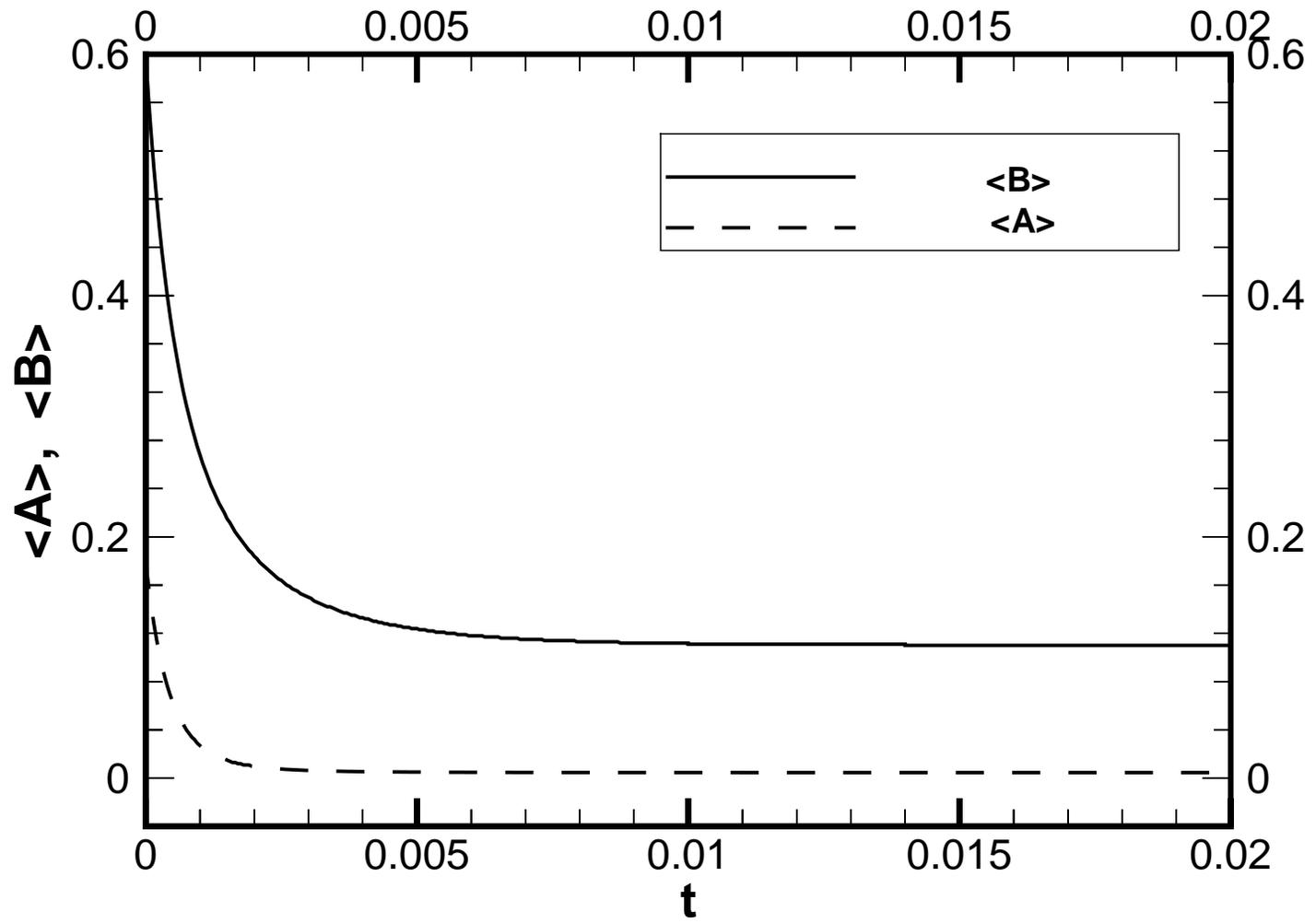
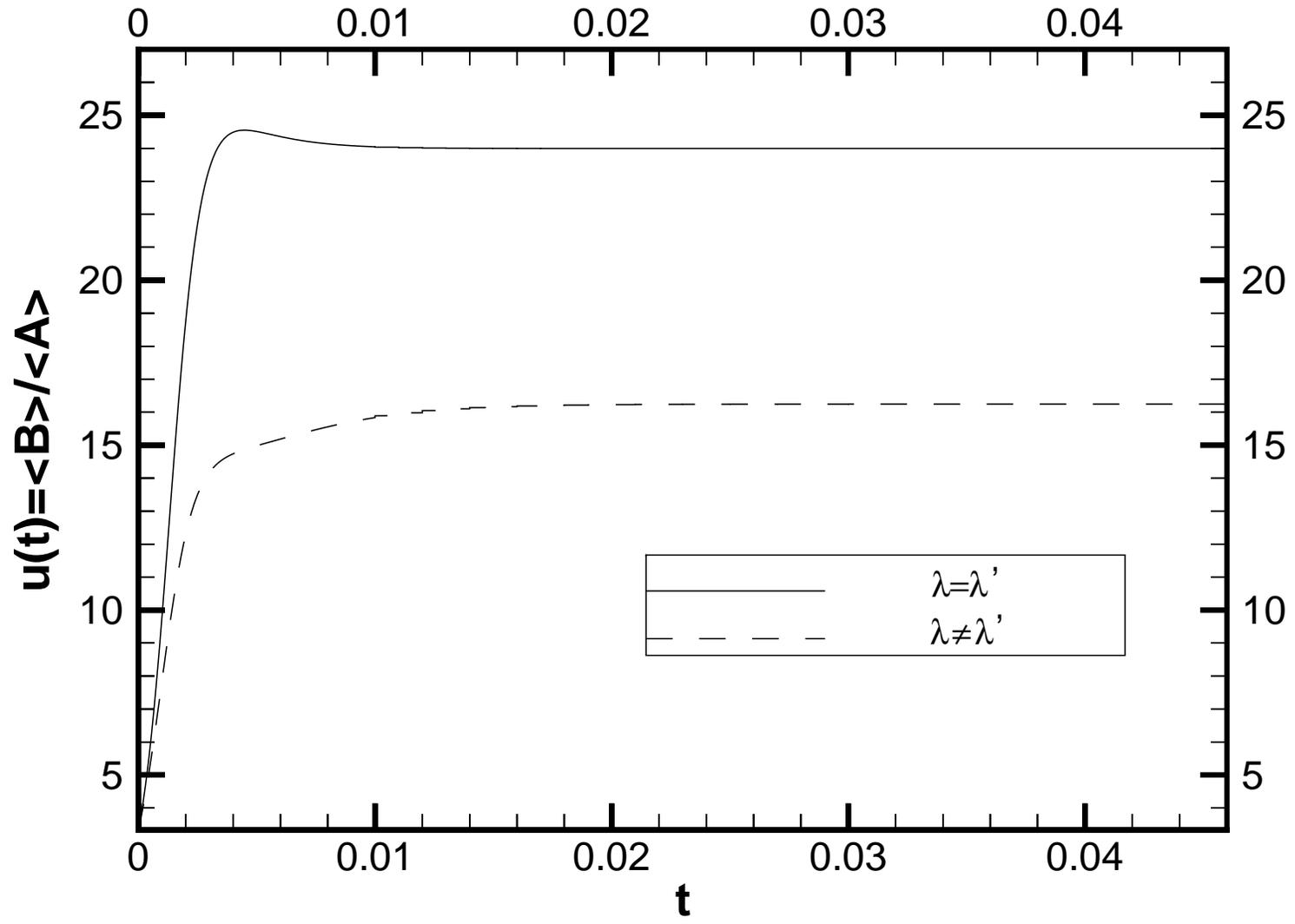
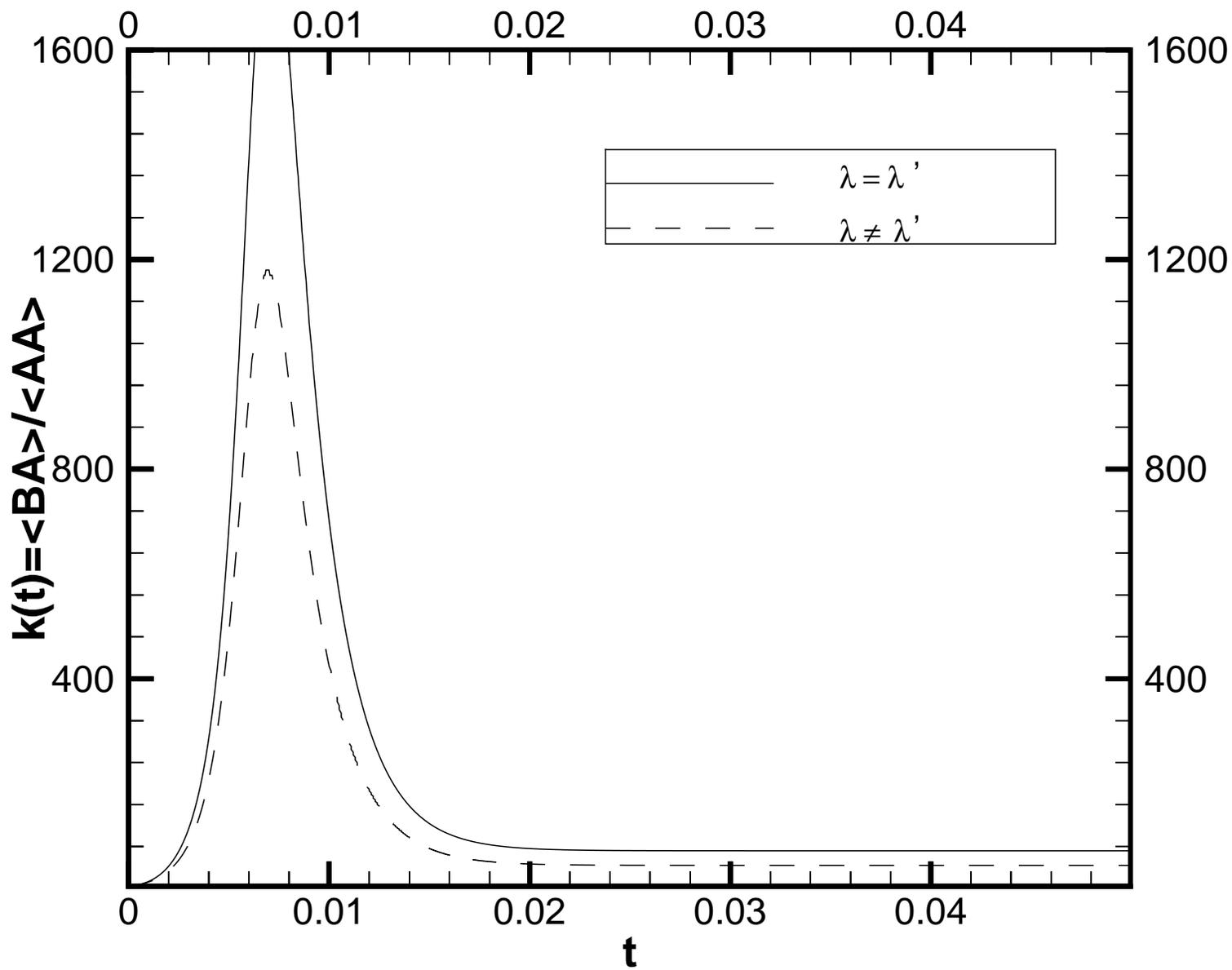
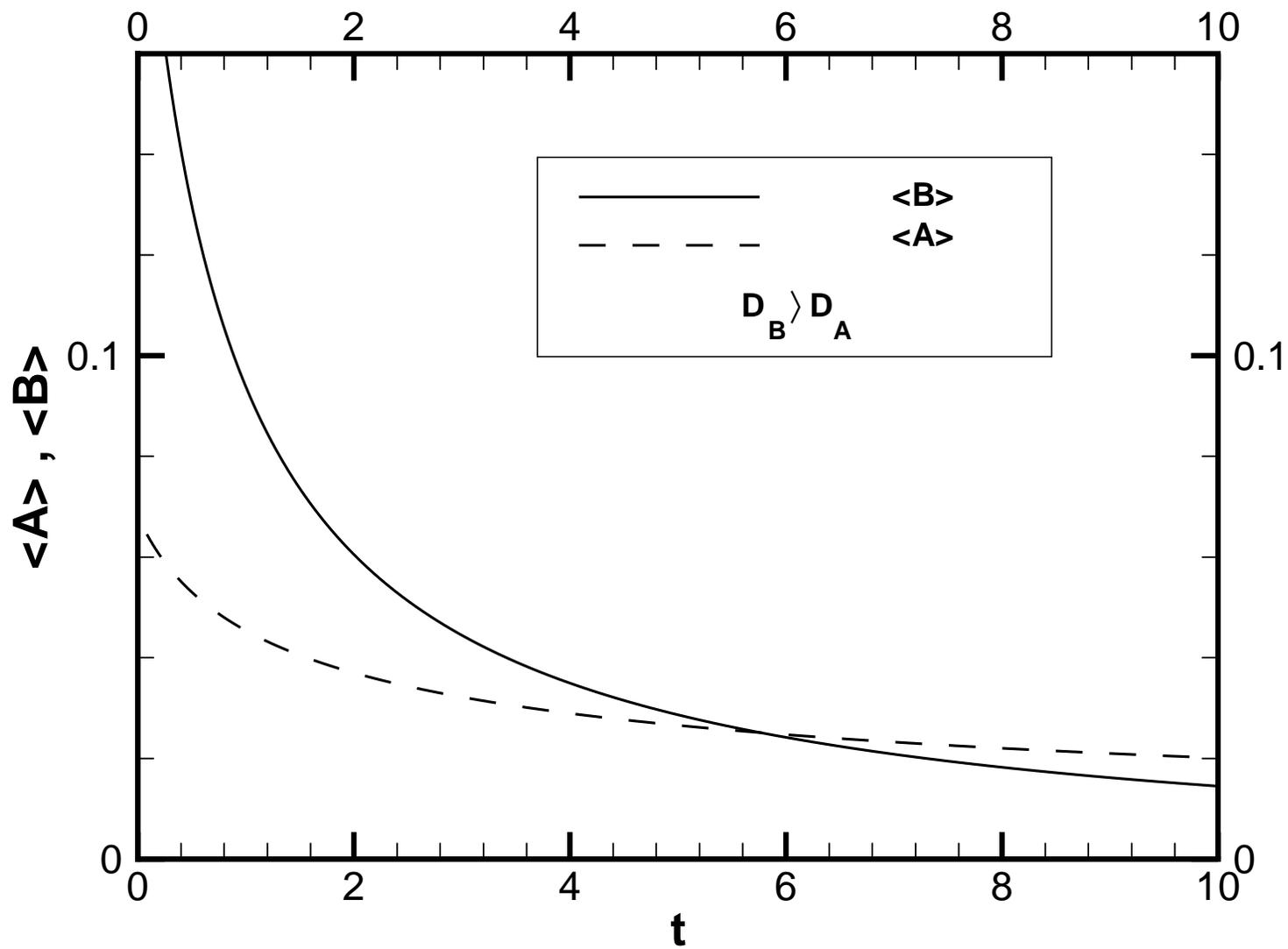
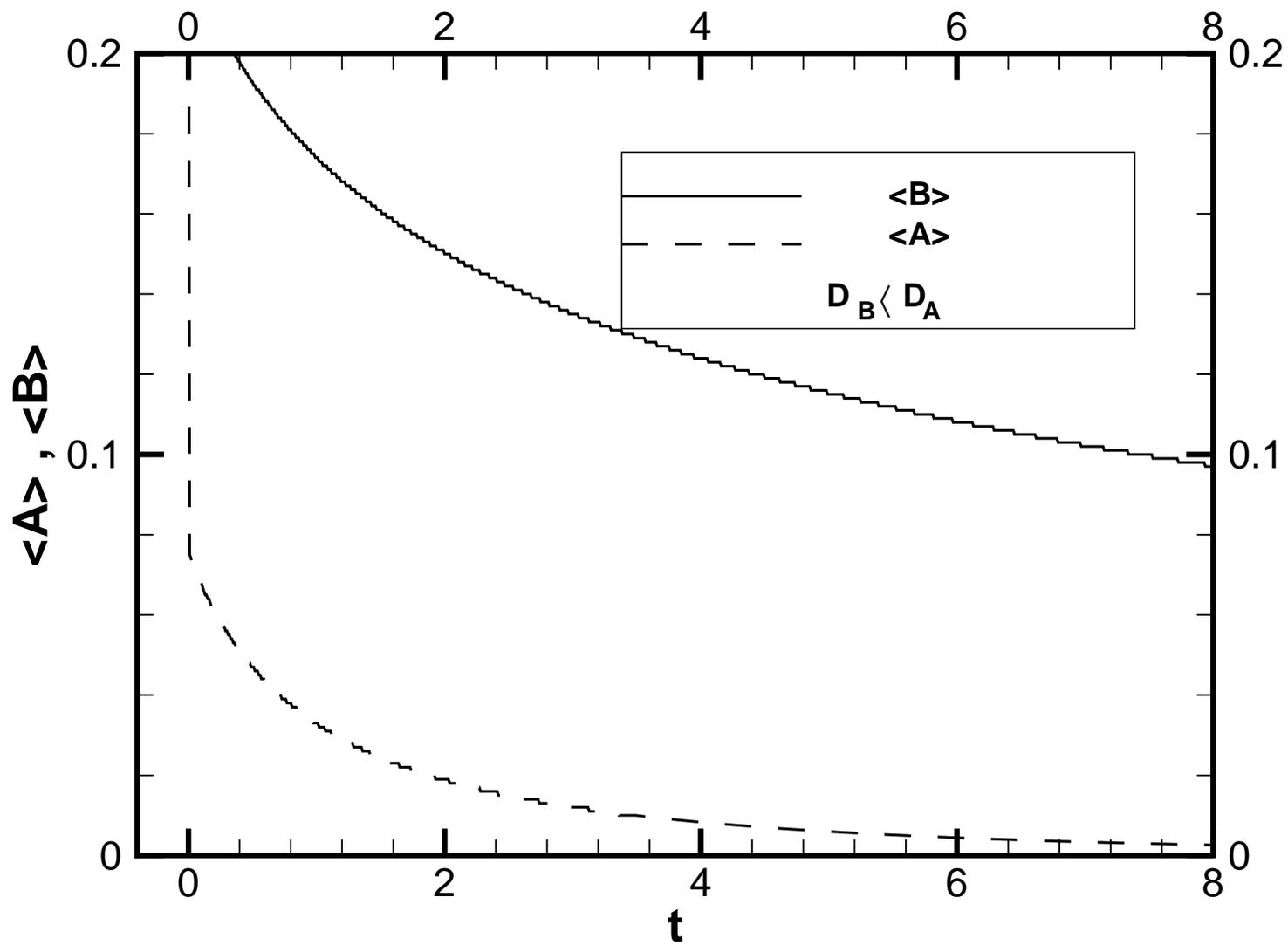
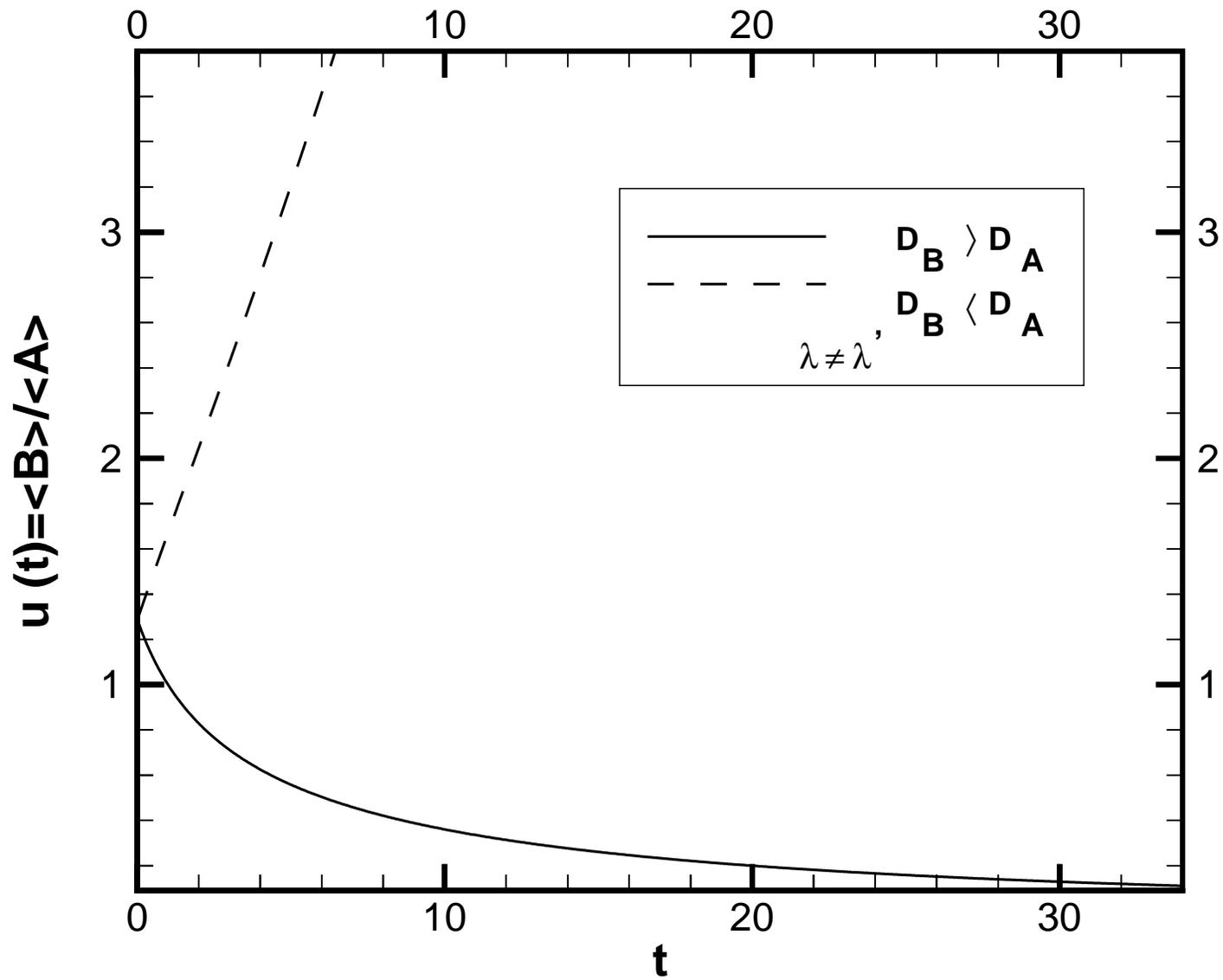
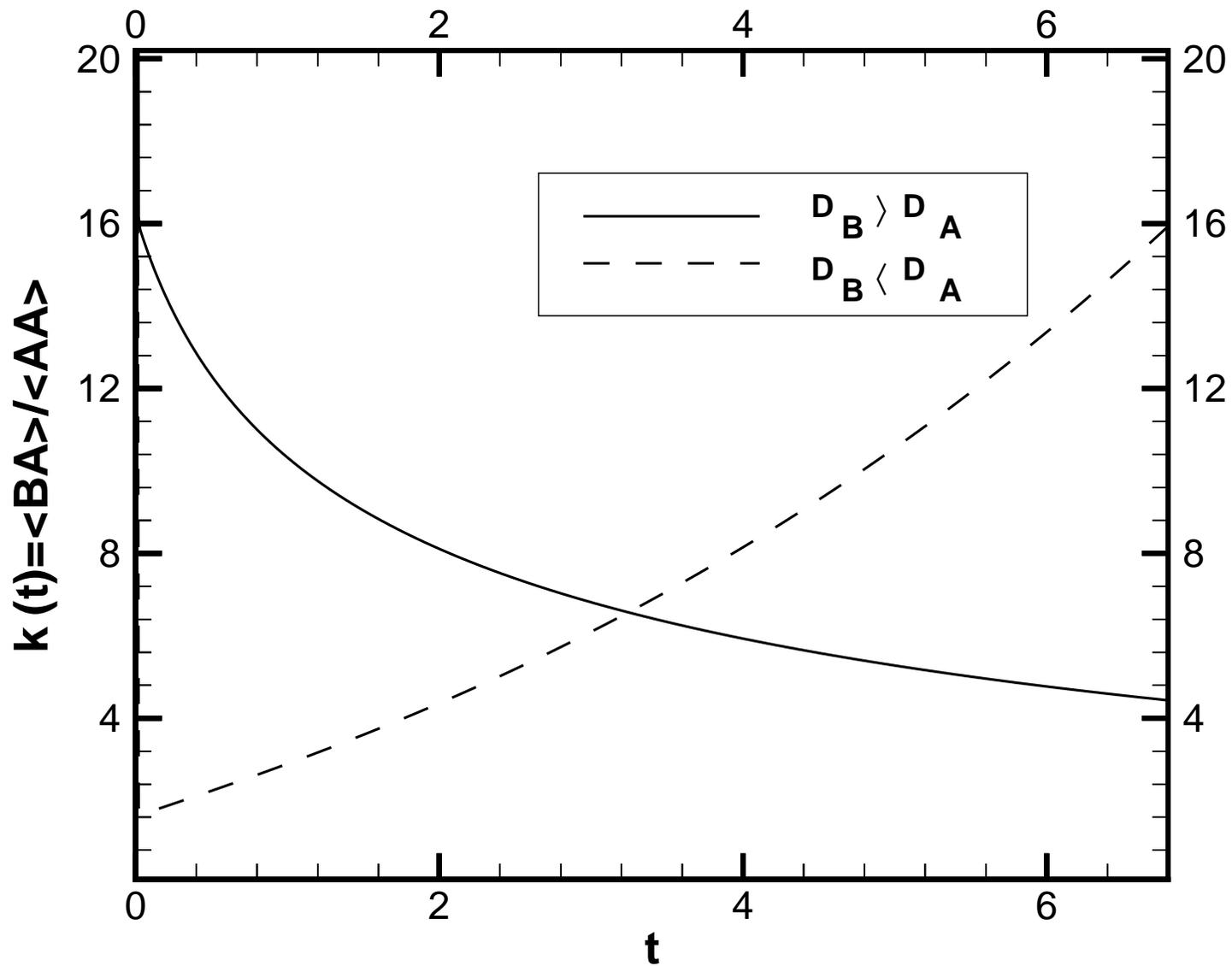
\end{document}